\documentclass[prb,aps,twocolumn,superscriptaddress]{revtex4-1}
\usepackage{graphicx,color}
\usepackage{amsthm}
\usepackage{amsfonts}
\usepackage{algorithmic}
\usepackage{enumerate}
\usepackage{latexsym}
\usepackage{amsmath}
\usepackage{amssymb}
\usepackage[colorlinks=true,citecolor=blue,linkcolor=blue]{hyperref}

\newcommand{\bc}{{\bf c}}

\def\avg#1{\left\langle#1\right\rangle}

\emergencystretch=\maxdimen
\begin{document}

\title{Metal-insulator transition and dominant $d+id$ pairing symmetry in twisted bilayer graphene}
\author{Wanying Chen}
\affiliation{Department of Physics, Beijing Normal University, Beijing 100875, China\\}
\author{Yonghuan Chu}
\affiliation{Department of Physics, Beijing Normal University, Beijing 100875, China\\}

\author{Tongyun Huang}
\affiliation{Department of Physics, Beijing Normal University, Beijing 100875, China\\}

\author{Tianxing Ma}
\email{txma@bnu.edu.cn}
\affiliation{Department of Physics, Beijing Normal University, Beijing 100875, China\\}

\begin{abstract}
Motivated by recent experimental studies that have found signatures of a correlated insulator phase
and tuning superconductivity in twisted bilayer graphene, we study the temperature-dependent conductivity, the spin correlation and the superconducting pairing correlation within a two-orbital Hubbard model on an emergent honeycomb lattice.
The evaluation of the temperature dependence of the conductivity demonstrates that there is a
metal-insulator transition, and the Mott phase at strong coupling is accompanied by antiferromagnetic order.
The electronic correlation drives a $d+id$ superconducting pairing to be dominant over a wide filling region.
All of the dc conductivity, the spin correlation and the superconductivity are suppressed as the interlayer coupling strength increases,
and the critical $U_c$ for the metal-insulator transition is also reduced.
Our intensive numerical results reveal that twisted bilayer graphene should be a uniquely tunable platform for exploring strongly correlated phenomena.
\end{abstract}
\maketitle

\section{Introduction}
In accordance with recent experiments on twisted bilayer graphene (TBG),
the arresting phenomena including novel phases, unconventional superconductivity and a Mott-like insulator behavior have been discovered in
an excellent two-dimensional system that demonstrates the importance of correlations effects\cite{Cao2018A,Cao2018B}. Most recently, a tuning superconductivity is induced by varying the interlayer spacing with hydrostatic pressure,
which sparked more intense interest in TBG, as it may be a uniquely tunable platform for exploring correlated states\cite{Yankowitz1059}.

With two layers of graphene twisted at a narrow range of particular magic angle,
the band structure of twisted bilayer graphene becomes nearly flat,
and as a result, the Fermi velocity decreases to zero in the vicinity of the Fermi energy.
Being interpreted as a correlated Mott insulator at half-filling\cite{Cao2018A},
this system is doped with a few additional charge carriers that then change the system from the initial insulator to a superconductor\cite{Cao2018B},
and shares a strikingly similar trend as that in doped cuprates\cite{Bednorz1986}, heavy-fermion\cite{Steglich1979}, iron-based\cite{Kamihara2008} and organic superconductors\cite{Jerome1980}. Consequently, it has been the subject of intense study since the discovery of high-temperature superconductors \cite{RevModPhys.84.1383}, which may shed light on several long-standing problems encompassing the understanding of unconventional superconductivity,
and may even prove to be a significant step in the search for room-temperature superconductors.

Substantial theoretical effort has gone into this field\cite{ROZHKOV20161,arXiv:1804.09674,PhysRevB.99.121407,PhysRevB.98.241407,PhysRevLett.121.217001,
zhang2018low,PhysRevB.98.195101,arXiv:1809.06772,HUANG2019310,PhysRevB.98.045103,PhysRevB.98.079901,Ray2018,Ochi2018,arXiv1809.00436,arXiv1805.06867,Guo2018Antiferro,
PhysRevX.8.031087,PhysRevX.8.031088,PhysRevLett.122.026801,Song2018All,
Ahn2018,PhysRevB.98.085435,Liu2018The,Carter2018Prediction,
Ming2018Moire,Wu2018Phonon,Wu2018Topological,PhysRevB.98.245103,Gu2019Antiferromagnetism,Wu2018Coupled,Lian2018The,
PhysRevB.98.241412,Xie2018On,PhysRevB.98.220504,Morna2019Flat,Chebrolu2019Flatbands,
Liao2019Is,Zhang2019Twisted,Laura2019Competing,Codecido2019Correlated,lfzhang2020},
and many possibilities of the exotic electronic structures further reflect the fact that TBG can be a realistic platform for various kinds of largely unknown physics.
However, it is difficult to identify an effective low-energy model to address the strong correlation effects in TBG,
as the moir\'{e} pattern in TBG with small twist angles requires a very large system size 
and more than 10, 000 atoms in one unit cell,
which makes the usual first-principles electronic structure calculations almost impossible. The origin of the insulating state and the possible pairing symmetry of the observed superconductivity are both highly debated\cite{arXiv:1804.09674,PhysRevB.99.121407,PhysRevB.98.241407,PhysRevLett.121.217001,zhang2018low,PhysRevB.98.195101,arXiv:1809.06772,HUANG2019310}
due to the weakness of the proposed effective model and the method to treat strong electronic correlation,
which are the major challenge in this active field. For example, some mean filed results argued that part of the experimental findings could well be understood
to be a $s$-wave superconducting state resulting from an attractive electron-electron interaction mediated by electron-phonon coupling\cite{PhysRevB.98.220504} and some recent works regarded the insulating state at half filling as a non-Mott picture but maybe some Kekul\'{e} valence-bond order\cite{PhysRevB.98.121406}, chiral spin-density wave order\cite{PhysRevLett.121.217001}, a nematic phases\cite{arxiv1804.03162}, or crystalline states\cite{doi:10.1021/acs.nanolett.8b02033}.

\begin{figure}[tbp]
\includegraphics[scale=0.22]{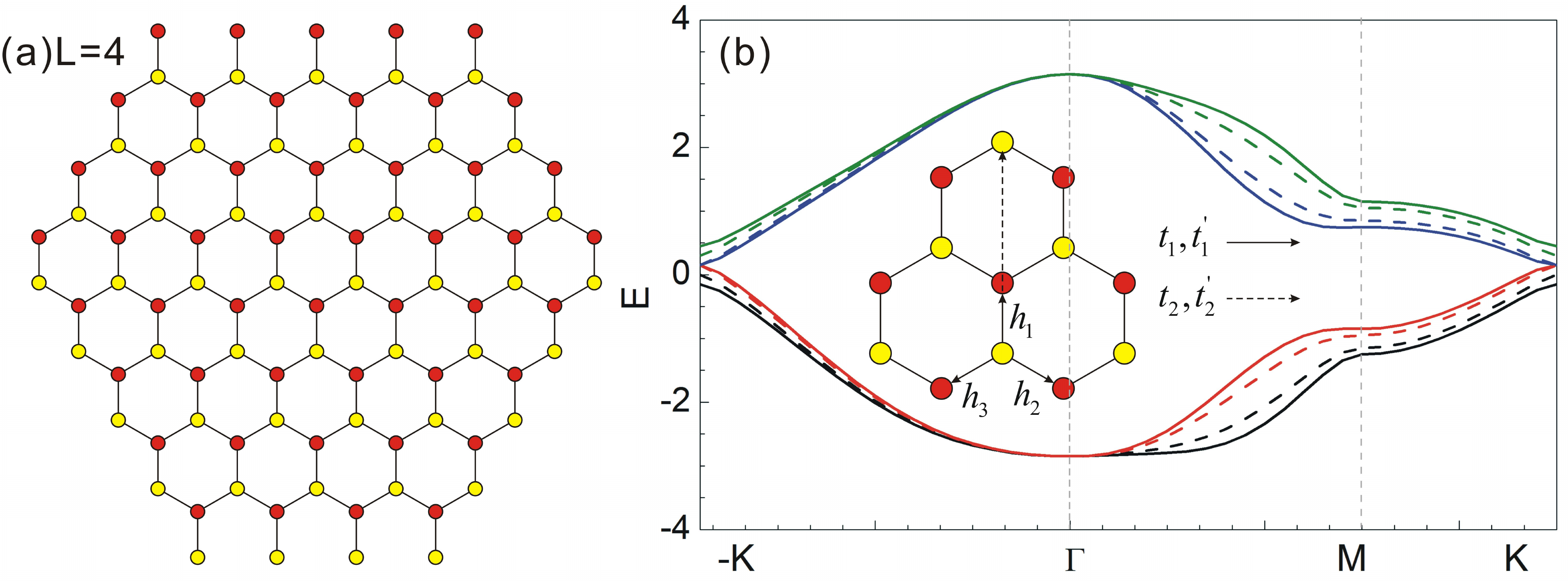}
\caption{(Color online) (a) The geometry of TBG for the effective model\cite{PhysRevB.98.045103,PhysRevB.98.079901} with $L=4$; (b) the band structure of the corresponding tight-binding model. Solid lines are $t'_1$=$t'_2$=0.10, and dotted lines are $t'_1$=$t'_2$=0.05. The insert shows the definition of hoppings.
} \label{structure_fu}
\end{figure}

Recently, a two-orbital Hubbard model, sketched in Fig.\ref{structure_fu},
constructed from Wannier orbitals that extend over the size of supercells,
is proposed to capture the electronic structure of narrow minibands and the effect of Coulomb interaction in
TBG, and the centers of these Wannier orbitals form an emergent honeycomb
lattice\cite{PhysRevB.98.045103,PhysRevB.98.079901}.
This model has been verified by explicit numerical calculations\cite{PhysRevX.8.031087,PhysRevX.8.031088}
and provides us an opportunity to explore the rich physics in TBG by the unbiased numerical method,
which is the most reliable way to establish the phase diagram in the system where the strong correlation effect dominates.
In this paper, we are aiming to identify the insulating state at half filling and the dominant pairing symmetry of the superconductivity with the proposed effective model.

The metal-insulator transition is one of the most fundamental
and profound physical phenomenon of quantum mechanics. However, our understanding
of interaction-driven metal-insulator transitions still remains
rather controversial because strongly correlated systems are
difficult to solve by using both analytical and numerical methods\cite{PhysRevX.6.011029}.
In this paper, with the two-orbital Hubbard model on an emergent honeycomb lattice, we perform a quantum Monte carlo study of the
metal-insulator transition and pairing interaction in TBG. The calculations of the current-current correlation function show that repulsion between electrons can significantly reduce the conductivity, and at low temperatures change the system from conducting behavior to insulating behavior. After performing a careful finite size scaling analysis, we demonstrate that the metal-insulator transition is associated with the presence of antiferromagnetic (AFM) long-range order.
Our intensive numerical simulations suggest that the state at half filling is a Mott-like insulating state with long range AFM order, and with a finite doping, the dominant pairing symmetry of the superconducting state is $d+id$-wave.


\section{Model and methods}
In the effective model, the tight-binding part $H_{tb}$ mainly contains the intralayer hoppings $H_0$, and $H_1$ and $H_2$ are also introduced into $H_{tb}$ to further break the SU(4) and SU(1) symmetries\cite{PhysRevB.98.045103,PhysRevB.98.079901}. The detailed $H_{tb}$ can be written as
\begin{eqnarray}
H_{tb}&=&H_0+H_1+H_2,\nonumber\\
H_{0}&=&\sum_{\langle ij\rangle}
t_1 [\bc^\dagger_{i}  \cdot\bc_{j} + h.c.]
+\sum_{\langle ij\rangle'} t_{2}[{\bc}^\dagger_{i}\cdot{\bc}_{j}+ H.c.],\nonumber\\
H_{1}&=&\sum_{\langle ij\rangle'}t_{2}'[ ({\bc}^\dagger_{i}\times{\bc}_{j})_{z}+ H.c.]\nonumber\\
&=&-i\sum_{\langle ij\rangle'}t_{2}' ({c}^\dagger_{i+}{c}_{j+}-{c}^\dagger_{i-}{c}_{j-})+ H.c.,\nonumber\\
H_{2}&=&\sum_{\langle ij\rangle} t'_1 [ \bc^\dagger_{i}  \cdot {\bm  ee}^{\parallel}_{ij}  \;
{\bm e}^{\parallel}_{ij} \cdot  \bc_{j} -\bc^\dagger_{i} \cdot {\bm e}^{\perp}_{ij}  \;
{\bm e}^{\perp}_{ij} \cdot  \bc_{j}+ H.c.],
\end{eqnarray}
where $\bc_{i}= (c_{i,x}, c_{i,y})^{\text{T}}$ with $c_{i,x(y)}$ annihilates
an electron with $p_{x(y)}$-orbital at site $i$. $ t_{1} $ and
$t_2$ are the hopping amplitude between nearest-neighbor(NN) and fifth-NN sites.
The sketch of hoppings is shown in the insert of Fig. \ref{structure_fu}(b).
The chiral basis $ c_{\pm}=(c_{x}\pm ic_{y})/\sqrt{2} $ is associated with $ p_{x}\pm ip_y $
orbitals. $\bm e^{\parallel,\perp}_{ij}$ denotes in-plane unit vectors in the direction
parallel and perpendicular to the NN bond $\langle ij\rangle$.
The on-site Coulomb interaction part is written as
\begin{eqnarray}
H_{U}=U\sum_{i,m}{\bf n}_{im\uparrow}{\bf n}_{im\downarrow}+V\sum_{i}{\bf n}_{ix}{\bf n}_{iy},
\end{eqnarray}
where $m$ is the $p_{x(y)}$ orbitals, and ${\bf n}_{im\sigma}=\bc^\dagger_{im\sigma} \cdot\bc_{im\sigma}$.
We have mainly used $V$=0 in this paper, except as explicitly noted elsewhere. In this model, $U_v(1)$ is kept and $C_6$ is completely ignored. Thus, in the Dirac points at charge neutrality are not symmetry-protected robust features\cite{PhysRevB.98.085435}. The unconventional superconductivity that we focused in this paper is away from the charge neutrality, and therefore, the controversial point of this effective model has no influence on our numerical calculations. Moreover, the electronic filling is controlled by the chemical potential, $H_{\mu}=\mu \sum_{i,m,\sigma}{\bf n}_{im\sigma}$. This model provides a theoretical framework for studying correlated electron phenomena in TBG. In our following simulations, the system we performed is sketched in Fig. \ref{structure_fu}(a) with periodic boundary conditions, and we take $t_1$ as the unit. 
With this effective model, the hopping parameters dictate bandwidth (at $\Gamma$ point), Dirac velocity (at $K$ point) and valley splitting (along $\Gamma-M$ lines). 
Such band properties can be calculated as functions of twist angle, and one could then compare them to extract hopping parameters.
We can then, choose the parameters that we find at the magic angle. The parameter we used is $t_1=1.0,t_1^{\prime}=0.1,t_2=0.025$, and $t_2^{\prime}=0.1$, which is taken from that of Refs.\cite{PhysRevB.98.045103,PhysRevB.98.079901}. With this set of parameters, one can see that the band structure in Fig. \ref{structure_fu}(b) is degenerate with finite $t_1^{\prime}$ including $\Gamma$ and $K$ points.

Our simulations are mostly performed on lattice of $L$=4,
and the total number of lattice sites is $N_s$=2$\times$2$\times$3$L^2$,
in which the first $2$ indicates two orbits,
and the second $2$ means two interpenetrating triangular sublattices with hexagonal shape such that it preserves most geometric symmetries of graphene\cite{PhysRevB.84.121410,PhysRevB.90.245114,doi:10.1063/1.3485059}. The number 3 means that each triangular sublattice is consistent of 3 square lattices with $L^2$ sites. The basic strategy of the finite temperature determinant quantum Monte Carlo (DQMC) method is to express
the partition function as a high-dimensional integral over a set of random auxiliary fields. The integral is then accomplished by
Monte Carlo techniques. In our simulations, 8 000 sweeps were used to equilibrate the system, and an
additional 30 000$\sim$ 240 000 sweeps were then made, each generating a measurement.
These measurements were divided into ten bins that provide the
basis of coarse-grain averages, and errors were estimated based on standard
deviations from the average. To assess our results and
their accuracy with respect to the infamous sign problem as the particle-hole
symmetry is broken, a very careful analysis on the average of sign is shown.

To explore the phase transitions between the metal and insulator behaviors, we compute the $T$-dependent DC conductivity, which is calculated from the wave vector $\textbf{q-}$ and imaginary time $\tau$-dependent current-current correlation function\cite{PhysRevLett.75.312} $\Lambda_{xx}(\textbf{q},\tau)$,
\begin{eqnarray}
\label{conductivity}
\sigma_{dc}(T)=\frac{\beta^2}{\pi}\Lambda_{xx}(\textbf{q}=0,\tau=\frac{\beta}{2})
\end{eqnarray}
where $\Lambda_{xx}(\textbf{q},\tau)=\left<\hat{j}_x(\textbf{q},\tau)\hat{j}_x(\textbf{-q},0)\right>$, $\beta=1/T$, $\hat{j}_x(\textbf{q},\tau)$ is the $(\textbf{q},\tau)$-dependent current operator in $x$ direction. Eq.\ref{conductivity} has been employed for metal-insulator transitions in the Hubbard model in many works, and it has already proved its validity\cite{PhysRevLett.75.312,PhysRevLett.83.4610,PhysRevLett.120.116601}.

To examine how the system evolves with the variation in
the magnetic order, we study the AFM spin structure factor
\begin{equation}
S_{AFM}=\frac{1}{N_s}\langle [\sum_{{mr}}({\hat{S}^{z}_{mar}}-{\hat{S}^{z}_{mbr}})]^{2}  \rangle,
\label{spin}
\end{equation}
which indicates the onset of long-range AFM order if $\lim_{N_s\rightarrow\infty}(S_{AFM}/N_s)>$0. Here, ${\hat{S}^{z}_{mar}}$(${\hat{S}^{z}_{mbr}}$) is the $z$ component spin operator on A (B) sublattice of orbit $m$. $S_{AFM}$ for different interactions
are calculated on lattices with $L=3,4,5,6$, and are extrapolated to the thermodynamic limit using polynomial functions in $1/\sqrt{N_s}$.

\begin{figure}[tbp]
\includegraphics[scale=0.39]{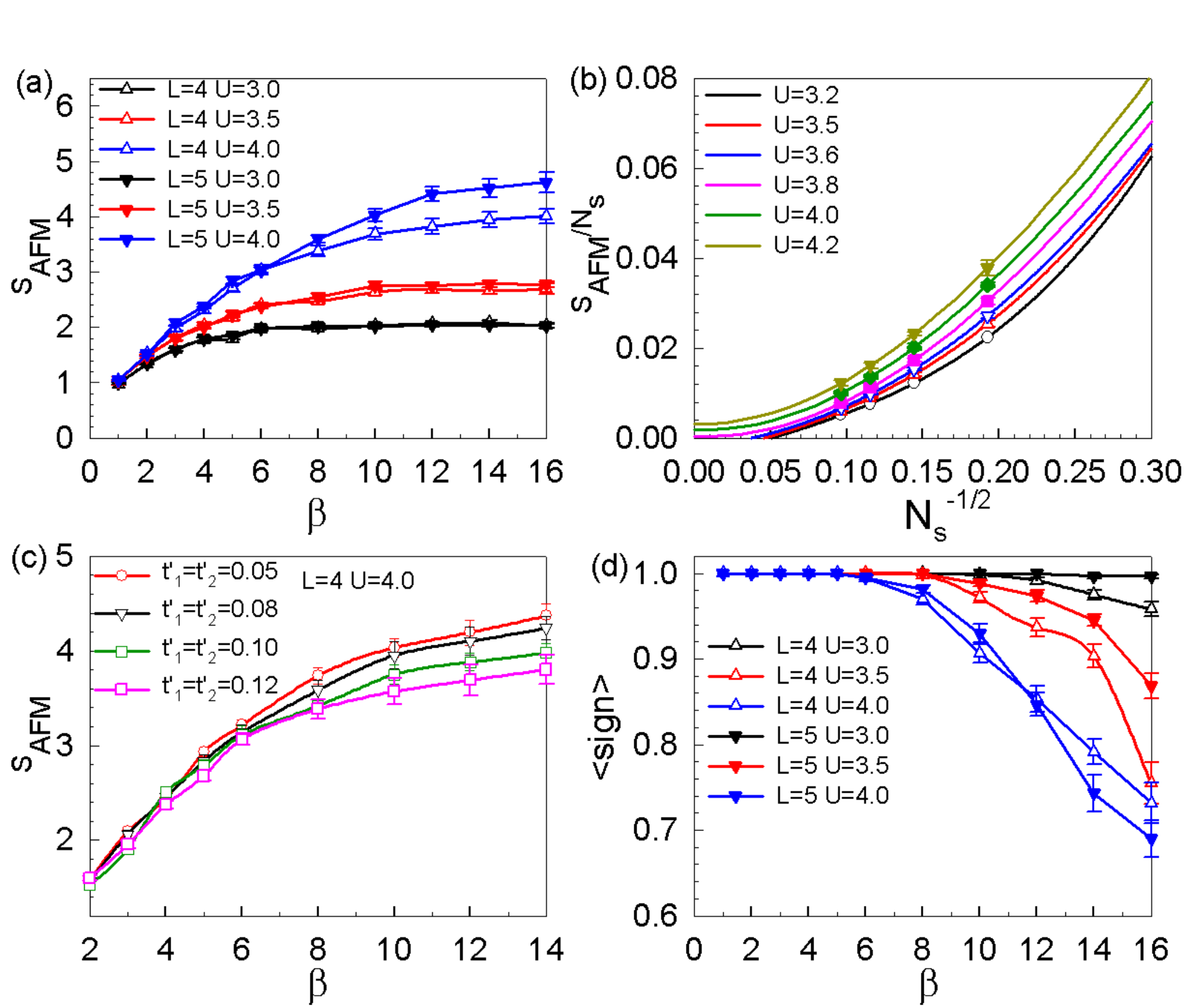}
\caption{(Color online) (a) The AFM spin structure factor $S_{AFM}$ depends on $\beta=1/T$ with different interaction strength and lattice size;
(b) the scaling behavior of the normalized AFM spin structure factor $S_{AFM}/N_{s}$ for different values of $U$ at $\beta=12$.
Solid lines are fit on the third-order polynomial in $1/\sqrt{N_s}$. (c)The AFM spin structure factor $S_{AFM}$ depends on $\beta=1/T$ at $L=4$ and $U=4$ for different $t'_1=t'_2$. The average of sign is shown in (d) for the simulation shown in (a).
} \label{Fig:AFM_Fu}
\end{figure}

\section{Results and discussion}
First, we examine the AFM spin structure factor behaviors versus the inverse temperature $\beta$ for the different lattice size, $L$, and interaction strength, $U$.
Here, we use $t_1=1.0,t_1^{\prime}=0.1,t_2=0.025$, and $t_2^{\prime}=0.1$.
In the following, we fix $t_1=1.0,t_2=0.025$, and may vary $t'_1=t'_2$ to explore the tunable physics.
From Fig. \ref{Fig:AFM_Fu}(a), we can see that the AFM spin structure factor increases as the lattice size increases slightly as $U\leq3.5$,
and the AFM spin structure factor clearly increases at $U=4.0$, which indicates that it has a potential to have a long range order as $U>3.5$.
To identify the accurate critical value of AFM long range order from Fig. \ref{Fig:AFM_Fu}(b), we further extrapolate the finite size results to the thermodynamic limit by using polynomial functions in $1/\sqrt{N_s}$,
and it can be seen from the figure that the AFM long range order starts to appear around $U \simeq 3.6\sim3.8$.
In Fig. \ref{Fig:AFM_Fu}(c), we also examine the behavior of the spin correlation as the interlay coupling strength varies, which shows that the spin correlation is suppressed as $t'_1=t'_2$ increases, which may be understood from the shown band width in Fig. \ref{structure_fu} (b). The increasing $t'_1=t'_2$ results in larger band width, and the effective interaction tends to be reduced and thus the spin correlation is suppressed.

For the finite temperature DQMC method,
the notorious sign problem prevents exact results for lower
temperature, higher interaction, or larger lattice for cases without particle-hole symmetry.
To examine the reliability of the present data shown in Fig.\ref{Fig:AFM_Fu}, we show the average of
sign in Fig. \ref{Fig:AFM_Fu}(c) and (d), which is
dependent on the different temperature $\beta$ at different interactions $U$ (a)
and different lattice sizes (b) with the Monte Carlo parameters of 30 000 times runs.
For the present results, our numerical results are reliable as one can see that the average of
the corresponding sign is mostly larger than 0.70 for the $U$ from
3.0 to 4.0 with 30 000 times measurements.
To obtain the same quality of data as $\avg{sign}\simeq1.0$,
much longer runs are necessary to compensate the fluctuations. Indeed, we can estimate that the runs need to be stretched by
a factor on the order of  $\avg{sign}^{-2}$\cite{PhysRevD.24.2278,SANTOS2003,PhysRevB.94.075106}.
In our simulations, some of the results are obtained with more than 240 000
times runs, and thus the results for the current parameters are
reliable.

\begin{figure}[tbp]
\includegraphics[scale=0.41]{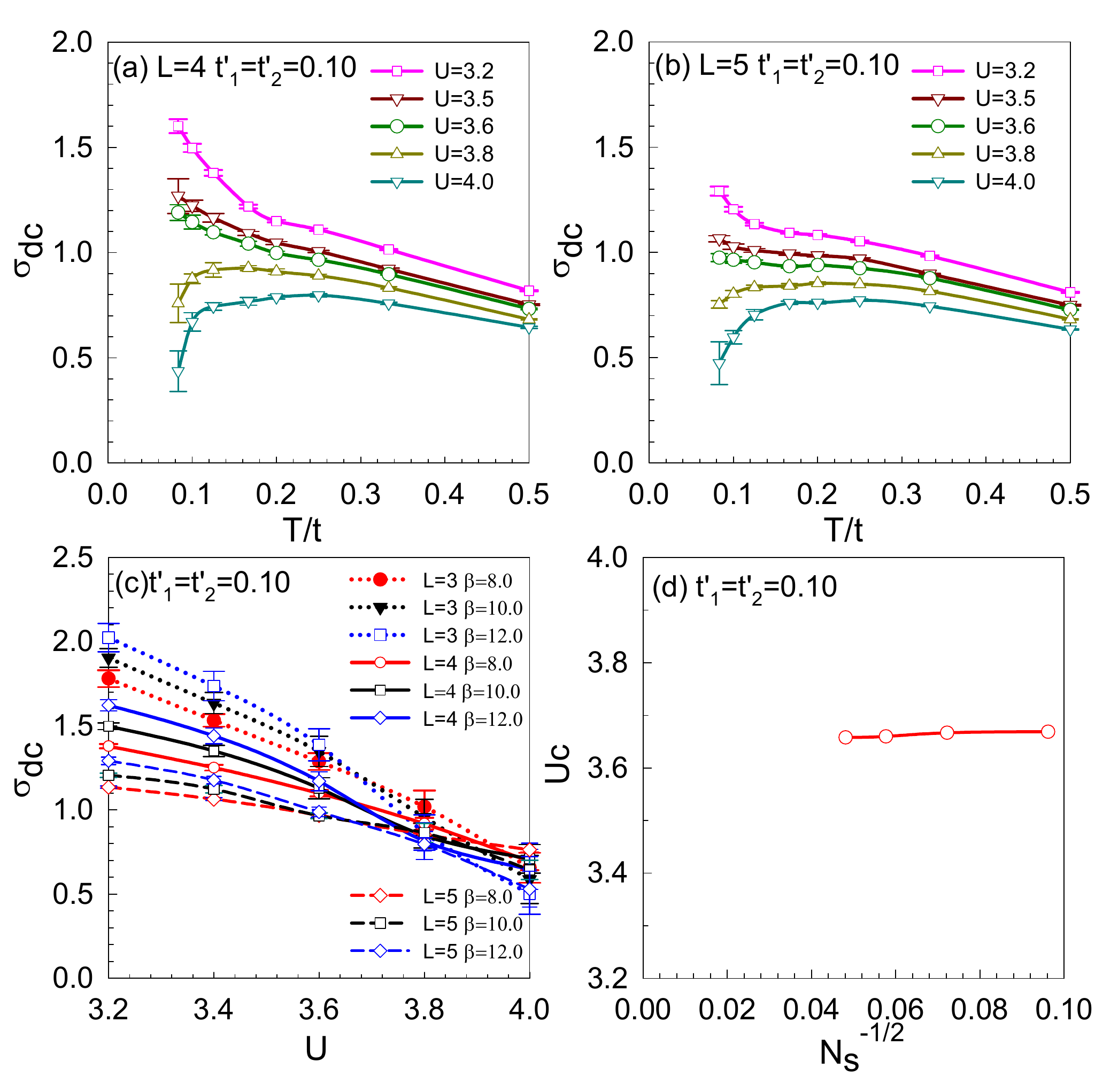}
\caption{(Color online) The dc conductivity, $\sigma_{dc}$, versus temperature, $T$, computed at various interaction strengths for (a) $L=4$ and (b) $L=5$.
 The conductivity $\sigma_{dc}$ versus the interaction $U$ for three different temperatures at $L$=3, 4 and 5 are shown in (c). The extracted $U_c$ for Mott-insulator transition with different lattices are shown in (d).
} \label{Fig:Dc_Fu}
\end{figure}

Second, we examine the temperature dependence dc conductivity $\sigma_{dc}(T)$ with $L=4,5$ across several interaction strengths,
as shown in Fig \ref{Fig:Dc_Fu} (a) and (b).
For $U<3.6$, $\sigma_{dc}$ diverges as the temperature $T$ decreases to the zero, while the conductivity curve is concave down and closed to zero with decreasing temperature for $U\geq3.8$. These behaviors of the $\sigma_{dc}$ curve in the low temperature region suggest that there is a metal-insulator transition at $U \simeq3.6\sim3.8$.
Moreover, the critical Hubbard interaction for AFM long range order is almost the same as that from semimetal to AFM insulator,
which reflects that there is no spin liquid phase in magic angle twisted bilayer graphene.
To further extract the critical interaction $U_c$ of metal-insulator transition and perform finite-size scaling analysis,
we plot the dc conductivity $\sigma_{dc}$ as a function of $U$ with three different temperatures in Fig.~\ref{Fig:Dc_Fu} (c) at $L=3,4,5$.
For each $L$, every three curves shown in Fig.~\ref{Fig:Dc_Fu} (c) intersect at one point, which reflects $U_c(L)$.
In the range of $U<U_{c}(L)$, the conductivity increases as the temperature decreases, which demonstrates the metallic phase.
The opposite situation emerges within the range of $U>U_{c}(L)$.
The conductivity $\sigma_{dc}$ values at higher temperature exceed those of lower temperature for the same $U$.
The system maintains an insulating phase.
The extracted $U_c$ for different lattice size have been shown in Fig.~\ref{Fig:Dc_Fu} (d),
and the critical interaction $U_c\simeq3.66$ is almost independent on the lattice size,
which ensure the $U_c$ for metal-insulator transition.

\begin{figure}[tbp]
\includegraphics[scale=0.41]{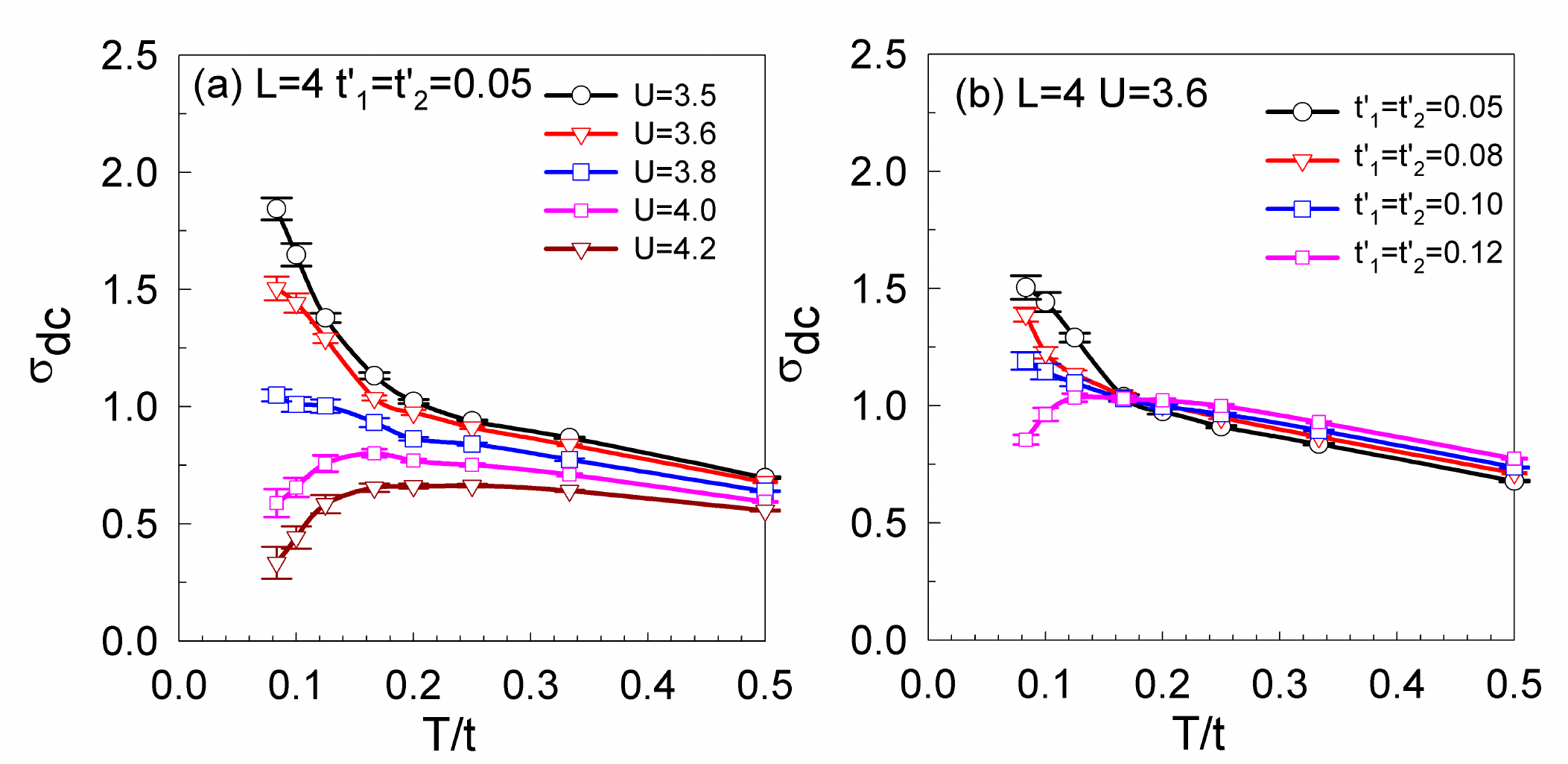}
\caption{(Color online) The dc conductivity, $\sigma_{dc}$, versus temperature, $T$, computed at
(a) various interaction strengths for $t'_1=t'_2=0.05$ and (b)at $U=3.6$ for different $t'_1=t'_2$.
} \label{Fig:t12p}
\end{figure}

It is interesting to ask what the fate of the correlated insulating phases is by varying the interlayer coupling $t'_1$ and $t'_2$, which may shed light on the relation between the superconducting phases and insulating phases.
In Fig.\ref{Fig:t12p} (a), the dc conductivity, $\sigma_{dc}$, versus temperature, $T$, computed at
various interaction strengths for $t'_1=t'_2=0.05$ is shown,
which indicates that the metal-insulating transition is in the region of $U_c= 3.8\sim 4.0$, which is slightly larger than that of $t'_1=t'_2=0.10$.
The dc conductivity for different interlayer coupling at $U=3.6$ is shown in Fig.\ref{Fig:t12p} (b), which
clearly shows that the dc conductivity reduces as the interlay coupling increases,
and that a smaller $U_c$ for metal-insulator transition is required for larger interlay coupling. Even the band width decreases as the $t'_1 = t'_2$ decreases, while the introduction of $t'_1$ and $t'_2$ may open a larger band gap, which favors
the insulating phases. The simulations that presented here is from an effective model, and one possible shortcoming here is that, it is not easy to
explore the effect of the twist angle or interlayer coupling on the metal-insulator transition.
It would be interesting to establish the correspondence with these physical parameters of the effective model, which allows,
 for instance, to discuss the effect of approaching the magic twist angle. 
 However, due to the limitation of the DQMC method of lattice size or the sign problem, it should be hard to extend that in current paper.

\begin{figure}[tbp]
\includegraphics[scale=0.41]{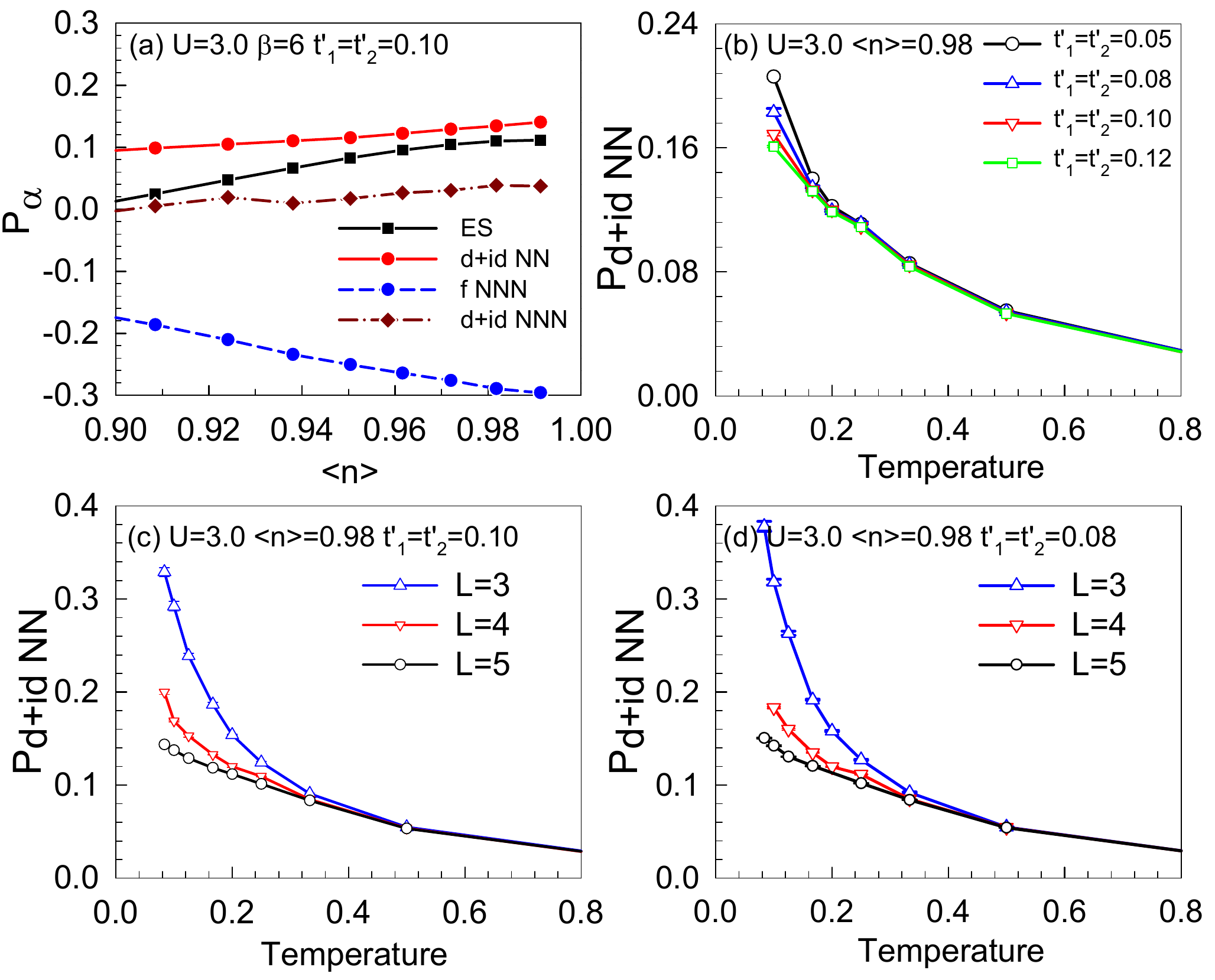}
\caption{(Color online) (a) The effective pairing interaction $P_{\alpha}$ with different pairing symmetry as a function of electronic fillings for $t'_1=t'_2=0.10$
and (b) the effective pairing interaction $P_{d+id NN}$ as a function of temperature for different $t'_1=t'_2$.
The effective pairing interaction $P_{d+id NN}$ as a function of temperature for $L=$ 3,4 and 5 are shown for $t'_1=t'_2=0.10$ (c) and $t'_1=t'_2=0.08$.
} \label{Fig:pair}
\end{figure}

To identify the pairing symmetry of superconducting state in TBG, we studied the effective pairing interaction with different pairing symmetries as
a function of the electronic fillings for $t'_1=t'_2=0.10$ in Fig.\ref{Fig:pair}.
Following previous literatures\cite{PhysRevB.39.839,PhysRevLett.110.107002,PhysRevB.40.506}, the effective pairing interaction ${\bf P_{\alpha}}$, is the difference between the pairing susceptibility $P_{\alpha}$ and the bubble contribution$\widetilde{P}_{\alpha}$, and it is defined as ${\bf P_{\alpha}}=P_{\alpha}-\widetilde{P}_{\alpha}$ with
\begin{equation}
P_{\alpha}=\frac{1}{N_s}\sum_{l,i,j}\int_{0}^{\beta }d\tau \langle \Delta
_{l,\alpha }^{\dagger }(i,\tau)\Delta _{l,\alpha }^{\phantom{\dagger}%
}(j,0)\rangle,
\label{Pa}
\end{equation}
where $\alpha$ stands for the pairing symmetry and the corresponding
order parameter reads
\begin{eqnarray}
\Delta _{l\alpha }^{\dagger }(i)\ =\sum_{\bf l}f_{\alpha}^{\dagger}
(\delta_{\bf l})(a_{{li}\uparrow }b_{{li+\delta_{\bf l}}\downarrow }-
a_{{li}\downarrow}b_{{li+\delta_{\bf l}}\uparrow })^{\dagger},
\label{Ca}
\end{eqnarray}
with $f_{\alpha}(\bf{\delta}_{\bf l})$ being the form factor of the pairing function.
To extract the intrinsic pairing interaction in the finite system, one should subtract from $P_{\alpha}$
its uncorrelated single-particle contribution $\widetilde{P}
_{\alpha}$, which is achieved by replacing $\langle
a_{{li}\downarrow }^{\dag }a_{{lj}\downarrow }b_{i+\delta_{\bf l}\uparrow}^{\dag}
b_{j+\delta_{\bf l'}\uparrow}\rangle $ in Eq. (\ref{Ca}) with $\langle a_{{i}\downarrow }^{\dag
}a_{{j}\downarrow }\rangle \langle b_{i+\delta_{\bf l}\uparrow }^{\dag }
b_{j+\delta_{\bf l'}\uparrow }\rangle $.

In Fig.\ref{Fig:pair}(a), it is shown that for the investigated filling region,
the pairing with $d+id$ symmetry dominates over pairings with other symmetry,
and agrees with our previous results on the original geometry\cite{HUANG2019310}.
The positive effective pairing interaction
indicates that there is indeed the possibility of electronic correlation driven superconductivity.
For the exact paring form, we refer readers to Ref.\cite{HUANG2019310}.
In Fig.\ref{Fig:pair} (b), the filling dependent effective pairing
interaction with $d+id$ symmetry is shown for different $t'_1=t'_2$.
The effective pairing interaction with $d+id$ symmetry, $P_{d+id NN}$ is
suppressed as $t'_1=t'_2$ increases. In Figs.\ref{Fig:pair}(c) and (d), we compare the effective pairing
interaction with $d+id$ symmetry on different lattices.
One can see that, the behavior that the effective pairing interaction increasing as the temperature decreases is robust,
indicating the possibility of superconducting transition in real systems.

\begin{figure}[tbp]
\includegraphics[scale=0.5]{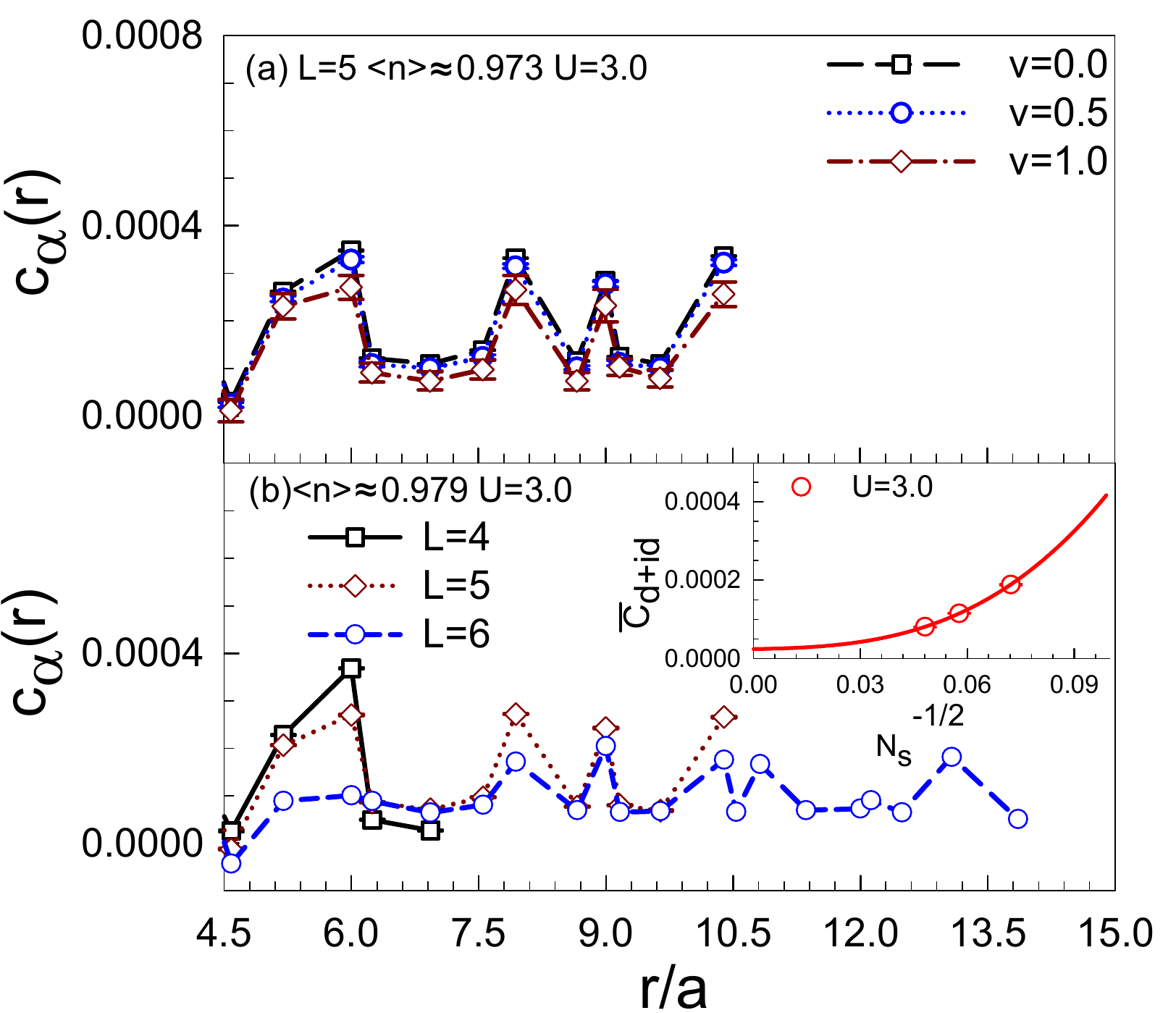}
\caption{(Color online) Pairing correlation $C_\alpha$ as a function of distance $r$ (a) for different inter-orbital Coulomb of $d+id$-wave with lattice of $L=5$ at $\left \langle n \right \rangle \approx 0.972$ and (b) with lattices of $L$=4,5, and 6 at $\left \langle n \right \rangle \approx 0.979$.
The inset of (b) shows the scaling analysis on the average of the long-range $d+id$ pairing correlation. Solid lines are fit on the third-order polynomial in $1/\sqrt{N_s}$.
} \label{Fig:PDr}
\end{figure}

In Fig.\ref{Fig:PDr}(a), we examine the effect of the inter-orbital Coulomb interaction on the ground state pairing correlation, $C_\alpha(r=\textbf{R}_i-\textbf{R}_j)=\Sigma_l\langle\Delta^{\dag}_{l\alpha}(i)\Delta_{l\alpha}(j)\rangle$, 
where one can see that the pairing correlation is slightly suppressed by the inter-orbital Coulomb interaction while the leading pairing symmetry does not change.
To learn more on the $d+id$ pairing correlation
in the thermodynamic limit, we examine the evolution of $C_{d+id}$
with increasing lattice size. In the inset of Fig.
\ref{Fig:PDr}(b), the average of long-range $d+id$ pairing
correlation, $\overline{C}_{d+id}$=$\frac{1}{\sqrt{N'}}\sum_{r>4a}C_{d+id}(r)$,
where $N'$ is the number of electronic pairs with $r>4a$, is plotted
as a function of $\frac{1}{\sqrt{N_s}}$ for $U$=$3.0$.
We observe that $\overline{C}_{d+id}$ has a finite positive value. This result
suggests the possible presence of long-range $d+id$ superconducting order in the parameter
regime investigated.

 In summary, within an effective two-orbit model for TBG, we study the spin correlation,
 the dc conductivity and the superconducting pairing interaction by using nonbiased DQMC method.
 At half filling, an antiferromagnetically ordered Mott insulator is proposed beyond a critical $U_c=3.6\sim 3.8$.
 By varying $t'_1 = t'_2$, we report a tunable metal-insulator transition and superconductivity in TBG,
 where the dc conductivity is suppressed as $t'_1 = t'_2$ increases, and a smaller $U_c$ for metal-insulator transition is required.
 With a finite doping, the pairing with $d+id$ symmetry dominates over the other pairing symmetries,
 and it could be suppressed by increasing the interlay coupling strength close to half filling.
 Our exact numerical results demonstrate that the TBG holds a very similar
 interaction driven phase diagram of doped cuprates and other high temperature superconductors,
 and the TBG opens up exciting opportunities to explore tunable correlated states in two-dimensional moir\'{e} superlattice heterostructures.

\noindent
\underline{\it Acknowledgement} ---
This work was supported by NSFC (Nos. 11774033 and 11974049) and Beijing Natural Science Foundation (No.1192011).
The numerical simulations in this work were performed at HSCC of
Beijing Normal University and Tianhe in Beijing Computational Science Research Center.
\bibliography{reference}

\end{document}